\documentclass[prl,twocolumn,floatfix,showpacs,showkeys]{revtex4} 

\usepackage[first,draft]{draftcopy}
\usepackage{latexsym,amssymb,amsmath,amsfonts}
\usepackage{graphicx}
\usepackage{bm}
\newcommand{\beq}{\begin{equation}}
\newcommand{\eeq}{\end{equation}}
\newcommand{\bc}{\begin{center}}
\newcommand{\ec}{\end{center}}
\newcommand{\eeqa}{\end{eqnarray}}
\newcommand{\beqa}{\begin{eqnarray}}
\newcommand{\no}{\noindent}
\newcommand{\pa}{\partial}

\newcommand{\ra}{\rightarrow}

\newcommand{\ga}{\gamma}

\newcommand{\de}{\delta}

\newcommand{\la}{\lambda}

\newcommand{\rh}{\rho}
\newcommand{\si}{\sigma}

\newcommand{\ph}{\phi}

\newcommand{\ps}{\psi}

\newcommand{\om}{\omega}
\newcommand{\ed}{\end{document} }

\begin{document}

\title{Spin flip probability of electron in a uniform magnetic field}
\author{Richard T. Hammond}

\email{rhammond@email.unc.edu }
\affiliation{Department of Physics\\
University of North Carolina at Chapel Hill\\
Chapel Hill, North Carolina and\\
Army Research Office\\
Research Triangle Park, North Carolina}

\date{\today}

\pacs{42.50.Ct}
\keywords{spin flip}

\begin{abstract}
The probability of a spin flip of an electron is calculated. It is assumed that the electron resides in a uniform magnetic field and interacts with an incoming electromagnetic pulse. The scattering matrix is constructed and the time needed to flip the spin is calculated.
\end{abstract}

\maketitle
   
\section{Introduction}
Controlling the spin state of an electron is becoming an increasingly important issue.
Spintronics continues to advance and managing the spin of the electrons is a vital concern.  Quantum information seeks long lived cubit states that allow reversible transfer and control.\cite{awschalom},\cite{zutic} Moreover, in generalized theories of gravity the spin gives rise to a new field, called torsion, which may someday be measured from spin flipping techniques.\cite{hammondreview},\cite{hammondprd}

A review of experiments involving spin flips in solids has been given,\cite{ortenberg} and more recently experimental investigations of nonlinear optical effects for spin polarized electrons in semiconductors has been undertaken.\cite{oleary}  Spin flip by a dc voltage in a quantum dot was also reported.\cite{smith}

For these reasons it is important to know the effect of an electromagnetic wave on the spin state. A classical investigation was undertaken by Gover,\cite{gover} who studied electron beams interacting with an electromagnetic pulse. While the main subject was spin resonant radiative emission, it was noted the spin flip is enhanced when coherent light is used. A related problem dates back to Brown,\cite{brown} who studied the geonium ``atom,'' which is a single particle placed in a Penning trap. In this case the magnetic field is not uniform.

In this Letter we compute the quantum mechanical probability of an electron spin flip induced by an electromagnetic pulse. In particular we consider an electron trapped in a uniform magnetic field, but otherwise free. Now we assume an electromagnetic pulse is incident on the electron and ask, what is the spin flip probability.

We start with the transition amplitude, defined by

\beq\label{ta}
S_{fi}=-ie\int d^4 x\overline\psi_f\ga^\si A_\si\psi_i
\eeq

\no where $f$ and $i$ denote final and initial states, 
$A_\si$ is the electromagnetic potential, and for now $\hbar=1=c$.

Normally the final and initial states are free particle wave functions, but that is not appropriate here. For one thing, from a philosophical view, it is not sensible to describe about the spin of an electron without a fiducial field by which one can measure the spin. In order to quantify the spin, there must a field, usually a magnetic field. Thus we consider the scattering of a bound state to another bound state, so the asymptotic wave functions should be those of an electron in a magnetic field, which is assumed to be constant.

The wave function for an electron trapped in a uniform magnetic field in the $z$ direction was found long ago and is given by\cite{huff}

\beq
\psi_n=C_fe^{-i(E_nt-p_x^nx-p_z^nz)}e^{-\xi^2/2}u_n
\eeq

\no where the momentum terms are eigenvalues, 

\beq
C_n^2=\frac{\sqrt{eB}(E_n+m)}{2L_xL_zE_n}
\eeq
and

\beq
\xi=\sqrt{eB}y -\frac{p_x}{\sqrt{eB}}
,\eeq

\no where the energy eigenvalues are

\beq
E_n^2=m^2+p_z^2+2neB
\eeq

\no  where $h_n=N_nH_n$, $H_n$ are the Hermite polynomials, $N_n=1/\sqrt{2^nn!\sqrt{\pi}}$, and by definition the Hermite polynomial with a negative subscript is zero. The $u_n$ are given below.

One precautionary note is given before the calculation is presented.
The cross section for spin flip is inherently small because the electromagnetic interaction mixes the ``large'' part of the Dirac spinor with the ''small part'' (whereas, for example, a chiral term mixes large with large). Thus other terms not involving spin flip are much larger, in general. These terms include momentum transfer effects, where the electron acquires a velocity (as in Compton scattering in lowest order). Thus, an electron at rest trapped in a uniform magnetic field can, besides spinflipping, acquire a net velocity from the interaction. However, if the electron is bound to a solid, then the electron will be held in place. However, in this case the the interaction Hamiltonian will differ than that used here, in that there is an additional potential to be included.
 
 In order to get an exact spin flip probability one must use the exact bound state wave functions and potential of a {\em bound} electron in an external field. However, such conditions are not known exactly, in general, and so the calculation used here may be taken as an estimate, provided the internal fields  are small compared to the external field. For example, in a typical paramagnetic salt the local magnetic field of the order of 50 mT, which is small compared to the 4 T field used here.

  Now we will calculate the probability of a spin flip starting with the scattering matrix
  
\beq
S_{fi}=\mbox{lim}_{t\ra \infty}<\psi_f|\psi_i>    
.\eeq  
  
\no As mentioned above, we are considering the scattering between two (quasi) bound states, but the analysis follows along the usual lines. We start with
the Dirac equation for a particle bound in a uniform magnetic field, with potential $A_\si$, that experiences an incoming electromagnetic wave with potential $\tilde A_\si$,

\beq
(i\ga^\si\pa_\si-e\ga^\si (A_\si+\tilde A_\si))\psi=m\psi
.\eeq

\no The propagator is defined according to

\beq
(i\ga^\si\pa_\si-e\ga^\si A_\si 
-m)\Delta =\de(x-y)
\eeq 

\no so that we have the general solution

\beq
\ps(x)=\ph(x)-i\int d^4y \Delta V(y)\psi(y)
\eeq

\no where $V= e\ga^\si \tilde A_\si$ and $\ph$ is the solution to the homogeneous equation. A full treatment of the propagator is not needed here since we are only interested in the first order effects, so that

\beq\label{ta2}
S_{fi}=\de_{fi} +\int d^4x \overline \psi_f V \ps_i
.\eeq
 The essential difference between this and (\ref{ta}) is that in (\ref{ta2}) the wave functions correspond to a free particle but in the above the functions are those of an electron in a magnetic field.

 Now we consider the spin flip such that the final state is spin up and the initial state is spin down. In this case we have,

\beq
u_f=\left(
\begin{array}{cccc}
h_{n-1} &\\
0 &\\
p_z^fh_{n-1}/(E_f+m) &\\
-\sqrt{2neB}h_n/(E_f+m) &\\
\end{array}\right)
\eeq
and

\beq
u_i=\left(
\begin{array}{cccc}
0 &\\
h_n &\\
-\sqrt{2neB}h_{n-1}/(E_i+m)  &\\
-p_z^ih_n/(E_i+m) &\\
\end{array}\right)
.\eeq

\no In the above all of the $h_n$ are $h_n(\xi)$, where $\xi$ is the dimensionless argument given above ($\xi=\sqrt{eB/\hbar c}y --p_xc/\sqrt{eB\hbar c}$ in cgs).

It is assumed that we have box normalization,
but only in a two dimensional box. This is so because $\psi$  is a function of $y$. Thus we may impose periodic boundary conditions in $x$ and $z$, but not in $y$. In essence, in order to obtain the Hermite polynomials, it is assumed the solution must vanish at infinity. Imposing different boundary conditions would yield solutions that differ than those given above.

The sides are given by $L_x$ and $L_z$.  It may be noted that the gauge freedom in the choice of the electromagnetic potential translates to a freedom in the choice of $y$ or $x$, or a combination of the two.

To begin the calculation, let us write (\ref{ta2}) as

\beqa\label{ta3}
S_{fi}=
-\frac{iC_fC_i}{L_0}\int d^4x 
e^{i(E_ft-p_x^fx-p_z^fz)}e^{-\xi^2}\nonumber \\
M_{fi}e^{-i(E_it-p_x^ix-p_z^iz)}
\eeqa
where
$L_0=\om/eE$ and
the matrix element is

\beq
M_{fi}=u_f^\dag\ga^0\ga^\si A_\si u_i
,\eeq

which becomes

\beqa\label{m}
M_{fi}=\ \ \ \ \ \ \ \ \ \ \ \ \ \\ \nonumber
A_0 \left(\frac{\sqrt{2}
   \sqrt{B (n+1) q} h_n h_{n+1}
   p_z^i}{(E_f+m)(E_i+m)}
   -\frac{\sqrt{2}
   \sqrt{B n q} h_{n-1} h_n
   p_z^f}{(E_f+m)(E_i+m)}\right)\\ \nonumber
+A_1
   \left(\frac{h_n^2
   p_z^f}{(E_f+m)}-\frac{h_n^2
   p_z^i}{(E_i+m)}\right)\\ \nonumber
+A_2 \left(\frac{i h_n^2
   p_z^i}{(E_i+m)}-\frac{i h_n^2
   p_z^f}{(E_f+m)}\right)\\ \nonumber
+A_3 \left(\frac{\sqrt{2} \sqrt{B
   (n+1) q} h_n
   h_{n+1}}{(E_f+m)}-\frac{\sqrt{2} \sqrt{B n q} h_{n-1}
   h_n}{(E_i+m)}\right)\\ \nonumber
.\eeqa   

Our main interest lies in spin flipping and not on translation. Therefore we focus here on the terms that are independent of the initial and final momentum (otherwise, if these eigenvalues are zero, we see that the matrix element vanishes). For this reason we take the electromagnetic potential to be of the form $A_\si=\{0,0,0,A_3\}$. 

We see from our choice of the potential, the matrix elements occur with Hermite polynomials with different indices. Due to the orthogonality we see that this term would vanish unless the potential is a function $y$. For these reasons, assuming the electromagnetic field is a Gaussian pulse, we take

\beq
E=E_0\cos(ky-\om t)e^{-(\frac{ky-\om t}{d})^2}
\eeq

\no where in the following we drop the subscript on the amplitude and $d$ is a dimensionless constant that essentially determines the number of wavelengths in the pulse. We assume that $d>>1$ so that the potential may be written as

\beq
A_3\approx-\frac E\om\sin(ky-\om t)e^{-(\frac{ky-\om t}{d})^2}\equiv\frac{E}\om{}f
,\eeq

\no which gives the electric field in the $x$ direction, the magnetic field in the $-z$ direction, and the wave propagating in the $y$ direction.

Using this in (\ref{ta3}) we can perform the $x$ and $z$ integrals yielding delta functions,
\beq
S_{fi}=-i\frac{C_iC_f}{L_0}XYa
\eeq

where $X\equiv 2\pi\de(p_x^f-p_x^i)$ and $Y\equiv
2\pi\de(p_z^f-p_z^i)$ and

\beq
a=\int dt dy e^{-\xi^2}fe^{i(E_f-E_i)t}
{\overline u}_f\ga^3A_3u_i
.\eeq
Now, we define $a_\pm=a_+ - a_-$ so that, after integrating over time,

\beqa
a_\pm=\frac{d\sqrt{\pi}}{2i\om}e^K\int dy e^{-\xi^2+b\xi} \nonumber\\
\sqrt{2eB}\left(
\sqrt{n_i}\frac{h_{n_f-1}h_{n_i-1}}{E_i+m}
-{n_f}\frac{h_{n_f}h_{n_i}}{E_f+m}
\right)
\eeqa

\no where $K=-(d\Delta^+/2\om)^2$, $\Delta^\pm=E_f-E_i \mp\om$, and $b=ikL_1(1+\Delta^+/\om)$.

For below, we define $L_1=1/\sqrt{eB}$ and $\phi=p_x/\sqrt{eB}$. To prepare for the final spatial integration we assume a photon is absorbed, so we are interested in $a_+$ which is simply called $a$ from here on, and write this as,

\beq
a=N_2L_1\int d\xi e^{-\xi^2 +b\xi} M_{fi}
\eeq

\no where 

\beq
N_2= \sqrt{\pi}\frac{dc}{2i\om }e^K
.\eeq

\no We make the substitution $\mu=\xi-b/2$, complete the square, and use the identity, 

\beq
H_n(x+y)=\sum_{k=0}^n
\left(\begin{array}{cc}
	n&\\
	k
\end{array}\hspace{-1em}\right)H_k(x)(2y)^{n-k}
\eeq

\no  where $\left(\begin{array}{cc}
	n&\\
	k
\end{array}\hspace{-1em}\right)
$ are the binomial coefficients,
so that

\beqa
a=N_3 \int d\mu\sum_{k=0}^{n_f}\sum_{k'=0}^{n_i}
 e^{-\mu^2}\\ \nonumber
\left(\begin{array}{cc}
	n_f&\\
	k
\end{array}\hspace{-1em}\right)
\left(\begin{array}{cc}
	n_i&\\
	k'
\end{array}\hspace{-1em}\right)
h_k(\xi)h_k'(\xi)b^{n_f-k}b^{n_i-k} -R
\eeqa

\no where $N_3=\sqrt{2eB\hbar c}N_2L_1e^{b^2/4}$ and $R$ is defined as the entire term preceding it with $n_f$ and $n_i$ replaced by $n_{f-1}$ and $n_{i-1}$.

The integrals may be performed without trouble since the $h_n$ are orthonormal with the weight function
$e^{-\xi^2}$. The result is

\beq
S_{fi}=-i\frac{C_iC_f}{L_0(E_f+m)}N_3\sqrt{n_f} XY\Sigma
-R
\eeq

\no where 

\beq
\Sigma\equiv
\sum_n^{n_f}b^{n_f+n_i-2n}
\left(\begin{array}{cc}
	n_f&\\
	n
\end{array}\hspace{-1em}\right)
\left(\begin{array}{cc}
	n_i&\\
	n
\end{array}\hspace{-1em}\right)
.\eeq

Now we define $P$ as

\beq
P=\int|S_{fi}|^2d\rh
\eeq

\no  where $\rho$ is density of states. Earlier we adopted two dimensional box normalization in the $x$ and $z$ directions, so the phase space of this volume is 

\beq
\rho=dp_xdp_z\frac{L_xL_z}{(2\pi)^2}
.\eeq
Performing the phase space integral we finally have,

\beq
P=\pi d^2\Sigma^2e^{2Q}\frac{E_i+m}{E_iE_f(E_f+m)}
\frac{BE^2c^3e^3}{\hbar \om^4}
\eeq
where $Q=b^2/4-d^2(\Delta^+)^2/4\om^2$.

If $P$ is less than one then we may interpret this as the probability that the electron suffers a spin flip. However, as defined $P$ may be greater than one, which physically corresponds to the fact that the electron will eventually flip back, and then again, and so on. In this case $P$ represents the number of times the particle flips. In many cases we are interested in investigating the parameters for which $P=1$. For example, let us calculate how long it takes for the particle to flip.

Suppose we consider the transition $n_i=0$, $n_f=1$.
Let us further assume that the electromagnetic field is tuned to the transition energy, and take $p_x=0$. With this the result reduces to $P=\pi d^2 E^2/B^2$, which may written in terms of the intensity of the electromagnetic wave, $I$ as (in cgs)

\beq
P= \frac{8\pi^2 d^2}{c B^2}I
.\eeq

Consider the question, how long must the pulse last in order to flip the spin.
Earlier we defined $d$ and the number of wavelengths in our pulse, so that $ d\la=ct$.  Now we solve for the time, $t=T$, such that $P=1$. The result is

\beq
T=\frac{\la B}{\pi}\sqrt{\frac{1}{8cI}}
.\eeq

One should note this result holds at resonance only, off resonance the scattering matrix decreases exponentially with time, and the corresponding spin flip time increases quickly.

This result seems to be compatible with a result of Hu et. al., who determine the  spin flip time of an electron to be greater than 0.1 $\mu$s. To emulate those results we consider a magnetic field of 4.0 T
in a field of 200 mW. To estimate the intensity we assume a distance of 30 cm and a standard 16 dB gain antenna, which gives $T=27$ $\mu$s. However, in that experiment, they considered a low density of electrons CdTe in quantum wells, so the comparison cannot be given too much weight.

Alternatively one may ask what intensity is required to flip an electron in a given time interval. Using the parameters given above, for example, it would take about 5 W cm$^{-2}$ to flip the electron in one microsecond.

In the above it was assumed that $d>>1$, which means there are many wavelengths in the pulse. It is natural to ask what the result for small $d$ is. To investigate the effects of a short pulse let us take the potential to be the Gaussian form of width $d$. This corresponds to a field that looks like a single wavelength of the order of $d$ in length. The calculation may be carried through as above, and the result is that the time needed to flip a pulse is reduced by a factor of $2d^4$.

In summary, the probability of the spin flip of an electron in a uniform magnetic field was calculated in the case that an electromagnetic pulse was incident on the electron. At resonance the time it takes to flip the spin was estimated.

\ed